# Reusing processes and documenting processes: toward an integrated framework

**Françoise Détienne**
Ergonomic Psychology Group
INRIA-Rocquencourt
Domaine de Voluceau,
BP 105, Rocquencourt
78153 Le Chesnay cedex
France
+33 1 39635522
Francoise.Detienne@inria.fr

**Jean-François Rouet**
Language and Communication Laboratory
95 Avenue du Recteur Pineau
86022 Poitiers cedex
France
+ 33 49885976
rouet@hermes.univ-poitiers.fr

**Jean-Marie Burkhardt**
Ergonomic Psychology Group
INRIA-Rocquencourt
Domaine de Voluceau,
BP 105, Rocquencourt
78153 Le Chesnay cedex
France
+33 1 39635196
Jean-Marie.Burkhardt@inria.fr

**Catherine Deleuze-Dordron**
INRIA-Grenoble
46 avenue Félix Viallet
38031 Grenoble cedex
France
Catherine.Deleuze-Dordron@imag.fr

**ABSTRACT**
This paper presents a cognitive typology of reuse processes, and a cognitive typology of documenting processes. Empirical studies on design with reuse and on software documenting provide evidence for a generalized cognitive model. First, these studies emphasize the cyclical nature of design: cycles of planning, writing and revising occur. Second, natural language documentation follows the hierarchy of cognitive entities manipulated during design. Similarly software reuse involves exploiting various types of knowledge depending on the phase of design in which reuse is involved. We suggest that these observations can be explained based on cognitive models of text processing: the van Dijk and Kintsch (1983) model of text comprehension, and the Hayes and Flower (1980) model of text production. Based on our generalized cognitive model, we suggest a framework for documenting reusable components.

**BACKGROUND AND OBJECTIVES**
Empirical studies have pointed out some critical aspects of the software design situation. Software design problems are known to be ill-structured (Guindon, 1990; Visser & Hoc, 1990). Specifications are lacking and must be inferred. There is a multiplicity of acceptable solutions for a design problem, and a multiplicity of constraints and criteria of evaluation for the solutions. No solution is optimal according to all criteria. Software design involves at least two knowledge domains: the problem domain and the programming domain.

One strategy to cope with this situation is to lean on previous designs. Empirical studies report that software design is rarely done from scratch. Designers reuse information about previous designs, both from external sources or internal sources (memory). An attempt to systematize and increase the efficiency of this strategy has been implemented in the software engineering approach referred to as "software reuse". In the literature on software reuse, reuse is often considered only as far as some external representation of a source is reused, for example by copying/modifying source code or using a specialisable component.

In this paper we will take a broader view of the reuse activity not limited to the external reuse view. In our view reuse refers to all situations in which information about an old solution is exploited during a new software development. Thus exploiting an old solution may refer to the reuse of an external representation of this solution but also to the use of related information inferred based on a source component even if this component is not actually reused. Furthermore software reuse may occur in any phases of software design. Sofware development is usually described as composed of three main phases: problem understanding, design, and coding. We will argue that exploiting an old solution may occur in any of these phases.

The software reuse issue is intrinsically linked to the documentation issue. Documenting programs properly is essential to support reuse. Our objective is to construct a generalized cognitive model which accounts for both reuse and documentation processes. Based on this model, we will suggest a framework for documenting reusable components.

**SOFTWARE REUSE AND DOCUMENTING: STATE-OF-THE-ART**
Software Reuse is currently one of the most active and creative research areas in Computer Science. This is mainly because software quality and productivity are assumed to be greatly increased by maximising the (re)use of (part of) prior design products instead of repeatedly designing from scratch (Gall, Jazayeri & Klösch, 1995; Krueger, 1989) . Supporting the reuse activity has become a big challenge in Software Engineering. The "large" definition of reuse which has been recently adopted by the software engineering community is:

"Software reuse is the reapplication of a variety of kind of knowledge about one system to another similar system in order to reduce the effort of development and maintenance of that other system. This reused knowledge includes artifacts such as domain knowledge, development experience, design decisions, architectural structures, requirements, design, code, documentation and so forth" (Biggerstaff & Perlis, 1989, p XV).

Even with this large definition, in practice, methods and techniques are implemented at the programming language level or code level and they aim to make code reuse safer. The main issue is to supply designers with "components" (e.g., code, specialisable components, class reuse by inheritance, specifications) and mechanisms to use these components into the current design. But in any cases, whatever the type of reusable components, reuse relies mostly on the existing code[1].

Currently, the predicted level of reuse has not been reached yet, due to technical, organisational and ergonomic factors (Biffl & Grechenig, 1993; Tracz, 1987) . No actual integration of reuse into process models and design methodologies has been proposed yet. Tools and environments are designed without taking reuse into account. Furthermore a common belief is that a "well-designed" component does not need to be provided with any further information. On the contrary, in practice, the need for software documentation is often emphasized as a critical factor in software reuse. One main issue is to properly document software components in order to support the reuse activity.

We believe that the main problem in software reuse is that it relies on the existing code. In this way, Green et al. (1992) insist on the fact that if the code is the essential element of the knowledge of the programmers, it "does not express all the knowledge that the programmer has or may wish to employ". These authors advocated the addition of a separate "description level" decoupled from the code; this decription level would be at the programmer's disposal for recording arbitrary attributes and relationships in a browsable form. This description level would be a kind of knowledge-base; it would record both transient facts and long-term facts about program components and program relationships.

However, as concerns software documenting, there is no agreed-upon methodology for documenting computer programs, and even less a general theory of how documenting relates to software design. Empirical studies (e.g., Rouet et al. 1995a) have shown that professional software designers may have different representations of what documenting is good for, even though they agree on general documentation guidelines (e.g., comments should support difficult algorithms). We have started to investigate the activity of software documentation with the general hypothesis that documenting is a natural component of skilled design. We claim that this approach is supported by evidence from a wide range of situations, which include documenting completed programs and documenting while designing a program.

## A GENERALIZED COGNITIVE MODEL

Empirical studies on design with reuse and on software documenting provide evidence for a generalized cognitive model. First, studies emphasize the cyclical nature of design: cycles of planning, writing and revising occur. Second, natural language documentation follows the hierarchy of cognitive entities manipulated during design. Similarly software reuse involves exploiting various kinds of knowledge depending on the phase of design in which reuse is involved.

We suggest that these observations can be explained based on the cognitive models of text processing: the van Dijk and Kintsch (1983) model of text comprehension, and the Hayes and Flower (1980) model of text production. These models account for (a) the hierarchy of entities manipulated during design-as-comprehension (textbase versus situation model) and (b) the cyclical sequence of planning-translation-revision in text production. In the two following sections we present a cognitive approach of documenting and of reuse. Results from empirical studies will be discussed in the text processing theoretical framework.

### A cognitive approach to documenting

In our perspective, documenting is an essential component of the design process. We claim that documenting serves two essential purposes: to assist the designer during problem-solving, and to improve the design product by making the program easier to communicate (i.e., easier to understand for another designer). Studies of program documentation have evidenced that designers output several types of comments. Some comments have to do with the surface structure of the program or textbase (e.g., paraphrases and syntactic explanations), others have to do with the problem being solved through the program or situation model (e.g., goal structure, justifications of design decisions). One dimension is of interest when analyzing the relation between documentation and software design: whether the designer is documenting an already completed program, or documenting as part of design. Empirical studies have shown interesting differences between those two situations.

*Documenting completed programs*
The task of documenting a program produced by someone else has been studied by Rouet et al. (1995b). The analysis of comments elicited several categories of information: paraphrases (comments that paraphrase program statements and do not include any new information), syntactic explanations, semantic explanations (about solutions being implemented), meta-comments (statements about commenting) and inferences from labels. When commenting simple procedures or simple programs, both intermediates and experts issued mostly semantic explanations (in this case, mostly low level functional information), then paraphrases. A general hypothesis was that comments reflect the designer's cognitive representation of the entity being commented. The results suggest that the representation constructed reflects (1) low level functional information close to elementary operations and (2) control flow information (paraphrases). This suggests that programmers have constructed a textbase representation. We refer here to the distinction between the textbase and the situation (or mental model) made in the van Dijk and Kintsch's model of text understanding (1983). The textbase represents what is said in the text and how it is said whereas the situation (or mental) model represents the situation referred to by the text. The structure of the textbase is isomorphic or homomorphic to the text structure whereas the structure of the situation model is not. The situation model is constructed on the basis of the textbase and inferences made by activating generic or episodic knowledge.

Another result from Rouet et al.'s study was that "structural" units, e.g., beginning of loops, were the most frequently commented. This suggests that programmers have constructed a textbase representation since the structure of the representation constructed reflects the structure of the program text (as defined by the control structure). To sum up, in the experimental task of documenting someone else's program, programmers tend to reason at the level of the textbase, i.e., they do not question the rationale for implementing a particular solution.

The same kind of results was found in another study on documentation (Riecken et al. 1991). Expert programmers documenting a program produced by someone else generated more comments detailing given instructions explicitly stated in the code rather than general domain information associated with the task. Subjects generated nearly twice as many detailed comments as abstract comments. These results suggest again that programmers in this kind of documentation task construct a textbase representation rather than a situation model. Furthermore, it was found that subjects located vertical spacing according to the program text structure, e.g., between routines. This last result also supports the textbase construction hypothesis since the constructed representation preserves the text structure.

The study of designers documenting someone else's program indicate that they may focus on a shallow level of representation: the program textbase, i.e. the surface and propositional structure.

*Documentation as part of design*
Software designers usually output a lot of natural language information as they design programs: design notes, temporary comments, or comments that are meant to remain with the final program (Davies, 1996; Henry et al. 1992). Even though the final goal of the activity is to write a list of code statements, natural language information is part of the designer's solution.

From this point of view, there is an interesting parallel between writing a program and writing a natural language text. One of the most influential cognitive model of text production is the one proposed by Hayes and Flower (1980). Hayes and Flower have defined three major phases in the writing process: Planning of the text structure as a function of domain knowledge (organizing) and communication purposes (goal setting); translating the text plan into a linguistic representation; and reviewing the text as a function of the writer's evaluation. One important feature of their model is that the overall process is cyclical rather than strictly linear. Moreover, the ordering of phases is not strictly predefined. Instead, the organization of writing phases is a function of the writer's strategy (Flower & Hayes, 1981). As a result, the writer generates a lot of intermediate results, e.g., planning notes, temporary text, additions and so forth. The writer also revises his or her drafts in order to improve clarity, coherence and/or to make the points stronger (Bereiter et al. 1988). Even though a first draft may be understandable, the polished product is usually better suited to the writer's purposes.

Software design also includes phases of planning, translation and revision (usually called problem solving or design, coding, revising). Gray and Anderson (1987) showed that such cycles occur in software development. Revising processes lead to changes characterised as stylistic, strategic and tactical. Stylistic changes involve revising *coding* whereas tactical and strategic changes involve revising *planning*.

Planning involves both retrieving problem-relevant knowledge and building up an abstract solution. Translating is equivalent to implementing the solution in a particular language. Finally, revising may include either modifying the implementation, the abstract solution, or even one's understanding of the problem structure. Some of the language generated during design will be later modified, some will be added, and some will be removed. Due to the particular purpose of computer programs -- to be used by both humans and machines -- the final product includes both natural and computer language. Following this approach, the status of documentation is not different from the code being generated. Instead, both documentation and code may be seen as a natural outcome of skilled design.

The empirical study of design with reuse provides good evidence that documenting is part of the design process (Rouet et al. 1995a). When asked to design with reuse of high-level components during the object-oriented design of a complex application, designers produce comments at all levels of abstractions: from the justification of high-level design decisions, to the description of implementation details. Moreover, designers have trouble narrowing the scope of their comments while designing. When asked to write only a particular category of information in a natural language field, they will often depart from the assigned category, and mix up different categories of information. We believe that this behavior reflects the cyclical-interactive nature of the design process. Both kinds of informations are tightly connected in the designers' representation, not as a function of their syntactic or semantic status, but as a function of the subproblem being solved at the moment. Some comments serve as an external memory, and they are meant to be deleted later (e.g., consequences of a design decision on another design decision to be made later on). Others -- e.g., justifications for particular solving decisions -- will stay as comprehension aids for the human reader. It is likely that the latter category contain information which connect the program surface structure (or "textbase") to the underlying solution structure (or "situation model"). However, the evidence for what comments are really helpful in program comprehension is still scarce.

**A cognitive approach to reuse**

Empirical work on software design with reuse has been conducted in the last years. Two contradictory results emerge from the literature. These studies tend to show, either that the effect of the reuse processes is an enrichment of the representations constructed during design (Burkhardt & Détienne, 1995a; Rosson & Carroll, 1993), e.g., by the inference of new constraints, new goals, or that the effect of the reuse processes is the lowering of the level of control of the activity during design, in particular by the use of test/debug strategies and comprehension avoidance strategies (Lange & Moher, 1989; Détienne, 1991). We defend the idea (Burkhardt & Détienne, 1995b; Détienne, 1996) that the former type of results concern reuse processes involved during the analysis and problem solving phases (which we call "reuse in planning") whereas the latter type of results concern reuse processes involved during the implementation phase (which we call "reuse in coding"). Our interpretation is that reuse in planning involves the construction of a situation model of the source whereas reuse in coding involves, but not necessarily, the construction of a textbase representation of the source.

*Reuse in planning*

The effect of the reuse processes may be an enrichment of the representations constructed during planning. Burkhardt and Détienne (1995a) show that evoking a reusable component may allow the addition of constraints, and new goals. Rosson and Carroll (1993) note that sometimes the borrowed code is not directly reusable itself but rather is used more as a functional specification. In another domain, architectural design, De Vries (1993) found that exploiting examples of old designs may allow the inference of new constraints for the new design and allow the constraints to be envisaged at a more abstract levels.

Several studies show that when a source component is evoked or retrieved during planning (as opposed to coding), information about the source situation from which the component comes from is searched or inferred. Burkhardt and Détienne (1995a) observed that this allowed programmers to infer solution goal structure, constraints, evaluation criteria or design rationales. Rosson and Carroll (1993) note the importance of knowledge about "example application" of a reusable class. In their field study, Rouet et al. (1995a) found that when selecting a reusable component in a library designers were looking for information on the application from which the component was extracted. This contextual information which seems to be highly important is rarely present in the documentation of components because software engineers generally believe that reusable components must be generic and application-independent.

In all these situations, it seems that reusing a component implies more than constructing a textbase representation of the source component itself. It implies constructing a situation model of the source application. In design with reuse, this situation model of the source allows the representation constructed for solving the problem on hand to be enriched and the search space to be enlarged.

*Reuse in coding*

Reuse results in the lowering of the level of control of the activity during the implementation phase. We refer to the hierarchy of levels of control developed by Rasmussen and Lind (1982). These authors distinguish between automatic activities, activities based on rules, and activities which involve high-level knowledge. The lowering of the level of control of the activity consists in switching from activities which involve high-level knowledge, e.g. problem solving activities, to activities based on rules and automatic activities, e.g. execution of procedures.

The use of the copy/edit style attests this effect. It reflects comprehension avoidance of the copied code and use of surface-level features to construct a representation of it[2] (Lange & Moher, 1989; Rosson & Carroll, 1993). The designers make "probable" modifications and rely heavily on the debugging tools to evaluate the code.

The analysis of what we called "new code reuse" episodes (Détienne, 1991) also shows that the level of control of the activity lowers. In these situations, the designer anticipates the reuse of a component in the same program while developing it. In new code reuse designers construct an operative representation of the source as well as a procedure for modifying the code of the source into targets. Designers then execute this procedure to develop targets. In this case also, there is a lowering of the level of control of the activity, which causes errors by propagation of source errors or by omission of changes. It was shown that the programmers constructed a representation of the source at various levels of abstraction. At the highest level, it represented the function of the source component whereas at the lowest level it represented the surface structure of the component, i.e. elementary operations. This suggests that a textbase representation of the source component is constructed.

**IMPLICATIONS**

We believe that both our studies on design with reuse and our studies on software documentation provide evidence for a generalized cognitive model, in which software design is seen as a particular instance of text processing. Based on this model we suggest a framework for documenting reusable components.

Our generalized cognitive model is based on the cognitive models of text processing: the van Dijk and Kintsch (1983) model of text comprehension, and the Hayes and Flower (1980) model of text production. Our model highlights two main aspects. The first aspect is that the nature of the design activity is cyclical. As in text production, the three main cognitive processes of planning, coding and revising occur in a cyclical way. Furthermore, depending on the process involved, the nature of knowledge manipulated varies. As in text processing, the nature of the representation constructed may be either a textbase representation or a situation model.

The second aspect of our model is that there is a parallelism between the nature of information reused depending of the design stage and the nature of documentation produced in various design stages. Studies of design with reuse have shown that reuse may occur during planning or coding. Reuse in planning involves the construction of a situation model of the source whereas reuse in coding involves, but not necessarily, the construction of a textbase representation of the source. Similarly, studies of documentation have indicated that documentation outputs -- e.g., inserted comments -- follow the hierarchy of entities manipulated during design: from abstract solution schemas or design rationales to low-level code statements.

Our contention that (a) design unfolds as a cyclical activity and (b) natural language documentation follows the hierarchy of cognitive entities manipulated during design has a potential implication for the design of design environments.

It has been shown that designers produce comments at all levels of abstractions: from justification of high-level design decisions, to the description of implementation details. Similarly software reuse involves exploiting various types of knowledge depending on the phase of design in which reuse is involved. We argue that a theoretical framework for documenting reusable components should be based both on our typology of reuse processes and on our typology of documenting processes. Clearly, documentation on justification of high level decisions linked to one or several components would be mostly useful when reuse is involved during planning (problem analysis and problem solving) whereas documentation on implementation details of one component would be useful when reuse is involved during coding. But we argue that both types of information are useful depending on the design phase.

The use of free annotations (Green et al. 1992) could be a way to implement this approach. We believe that the tools provided to store and edit documents should not assign a particular status to the information being stored. Chances are that during the design phases the designer will have to store information at various levels at the same location. On the other hand, the tools should include some type of a documentation checker, so that the designer can trim documentation after the design is completed, and remove the pieces of information that are no longer relevant.

Finally, we should point out a limitation of this approach. It is common to think that a design has a rationale or that the design space, i.e., the space of possible designs, can be analyzed by designers. In our approach, this information could be belong to the documentation of a reusable component. However a limitation of this approach on documentation, which also applies more generally to the design rationale approach, has been discussed recently by Karsenty (1996). This author made an empirical evaluation of design rationale documents. He found that designers having to evaluate a previous design for reusing it asked design rationales questions not answered by the design rationales documentation provided with the previous design. These kinds of questions, not handled by the designers of the source, could not be avoided and in fact it is likely that the number of this class of questions will increase as the time goes by and designer's culture changes. Karsenty speculates that long term reuse of design rationales should be more problematic that short-term reuse.

## ACKNOWLEDGEMENTS
This research was partially sponsored by the European Esprit III SCALE 6334 project (System Composition and Large Grain Component Reuse Support). Thanks to Denis Alamargot for his insightful comments on the parallel between software design and text production.

---

[1] An alternative approach, which seems promising, is based on analogical reasoning in the reuse of software specification (Lung et al. 1995; Maiden, 1991).

---

[2] These observations have been made in the object-oriented design. It is worth noting that the copy/edit style does not conform to the style of reuse encouraged for OOP, i.e., reuse by inheritance.